\documentclass[letter]{sig-alternate-2013}

\newfont{\mycrnotice}{ptmr8t at 7pt}
\newfont{\myconfname}{ptmri8t at 7pt}
\let\crnotice\mycrnotice%
\let\confname\myconfname%

\usepackage[english]{babel}
\usepackage[utf8]{inputenc}

\usepackage{subfigure}
\usepackage{microtype}

\usepackage{booktabs}
\usepackage{multirow}

\permission{}
\copyrightetc{}

\begin{document}

\title{Information Centric Networking in the IoT: \\Experiments with NDN in the Wild}
%
%
%
%
%

\numberofauthors{5} 
%
\author{
%
%
\alignauthor Emmanuel Baccelli \\
       \affaddr{INRIA}\\
       {\normalsize emmanual.baccelli@inria.fr}
\alignauthor Christian Mehlis\\
       \affaddr{Freie Universit\"at Berlin}\\
       {\normalsize mehlis@inf.fu-berlin.de}
\alignauthor Oliver Hahm\\
       \affaddr{INRIA}\\
       {\normalsize oliver.hahm@inria.fr}\\
\and
\alignauthor Thomas C. Schmidt\\
       \affaddr{HAW Hamburg}\\
       {\normalsize t.schmidt@ieee.org}
\alignauthor Matthias W\"ahlisch\\
       \affaddr{Freie Universit\"at Berlin}\\
       {\normalsize m.waehlisch@fu-berlin.de}
}

\maketitle

\begin{abstract}
This paper explores the feasibility, advantages, and challenges of an
ICN-based approach in the Internet of Things. We report on the first NDN
experiments in a life-size IoT deployment, spread over tens of rooms on
several floors of a building. Based on the insights gained with these
experiments, the paper analyses the shortcomings of CCN applied to IoT.
Several interoperable CCN enhancements are then proposed and evaluated.
We significantly decreased control traffic (i.e., interest messages) and
leverage data path and caching to match IoT requirements in terms of
energy and bandwidth constraints. Our optimizations increase content
availability in case of IoT nodes with intermittent activity. This paper
also provides the first experimental comparison of CCN with the common
IoT standards 6LoWPAN/RPL/UDP.
\end{abstract}


\section{Introduction}

The Internet is currently evolving in several directions. One path leads beyond end-to-end streams with Peer-to-Peer, CDNs and now ICN \cite{ICN-survey}. Endpoints in these information access models try to access named content, without direct mapping to a transport layer session at the (single) origin. The other evolves beyond traditional user terminal vs. router dichotomy: machine-to-machine (M2M) communications do not involve human source or destination, and interconnected machines include billions of cheap tiny communicating objects which play both the roles of host and router in spontaneous (wireless) networks, i.e., the Internet of Things \cite{atzori2010-IoT-Survey}. In this dual context, this paper explores the feasibility, advantages and challenges of an ICN-based approach in the Internet of Things.

\subsection{The Next Billion of Connected Machines}

The next billions of interconnected machines are expected to consist in a variety of heterogeneous devices, ranging from wireless sensors to actuators, wearables, Radio-Frequency IDentification (RFID) tags, smart home appliances and many other types of machines that were typically not inter-networked so far. Connecting these devices to the global realm has been coined the Internet of Things (IoT). It is expected to profoundly transform our environment.


Most IoT devices will be very limited in terms of memory, CPU, or power capacities (from small batteries). The term \emph{constrained devices} \cite{rfc7228} was recently introduced to define a category of connected devices that are resource-challenged compared to PCs, smartphones or laptops. Constraints include (i) orders of magnitude less power consumption measured in mWatt instead of Watt, (ii) orders of magnitude less computational power measured in MegaFLOPS instead of GigaFLOPS, and (iii) orders of magnitude less memory measured in Kilobytes instead of Gigabytes. For cost reasons, and due to the specific nature of the envisioned (massive) deployments of IoT devices, such constraints are expected to remain the norm in this domain, in the foreseeable future~\cite{Quartz-IoT-Moore}.

The sheer numbers and a lack of user interfaces make interconnecting IoT devices a challenge. Different approaches have been designed which leverage both traditional, infrastructure-based network paradigms, and spontaneous wireless network paradigms \cite{cordero2013-SIGCOMM-eBOOK}. They allow for device autoconfiguration and dynamic self-organization to relay data towards destination -- even without the help of infrastructure and pre-provisioned access points. Current approaches fall into two categories: silo approaches such as Zigbee \cite{alliance2012zigbee}, and approaches based on open standards, protocol stacks, such as IPv6 with 6LoWPAN \cite{rfc4944} and RPL \cite{rfc6550}. In the long run, one can expect that for the same reasons that led TCP/IP to prevail, an approach based on open standards and on a layered protocol stack will establish in the IoT. In the following, we will consider 6LoWPAN/IPv6/RPL as the reference networking solution for constrained devices in the IoT, with which ICN should measure up. 

\subsection{Why ICN for the Internet of Things?}

Data in information-centric networking is delocalized and need not be retrieved via an end-to-end transport stream. Instead, hop-wise replication and in-network caching facilitate information dissemination in the IoT, and relax the demand for continued connectivity. Such perspectives, based on ICN, were recently mentioned as a potential alternative networking solution for the IoT \cite{ghodsi2012-Dagstuhl-report}.

More specifically, common communication patterns of the IoT such as  content retrieval `upon request' and `scheduled' content updates are easily accommodated by ICN and may noticeably benefit from cache-assisted, hop-by-hop replication. The prevalent task of data fusion in the IoT may be implemented by augmented replication logic in a lightweight fashion. The combination of these mechanisms may save energy and radio resources, increase availability, and well reduce complexity. Most strikingly, ICN does reduce network layers and -- in an optimized version -- may subsume network, transport, and elementary application logic. Thus an ICN approach in the IoT might (i) offer opportunities to efficiently factorize functionalities  e.g., caching and buffering for error control (ii) drastically reduce the complexity of autoconfiguration mechanisms compared to an approach based on a layered protocol stack, and (iii) achieve a smaller memory footprint compared to 6LoWPAN/IPv6/RPL.  

However, a number of challenges should also be noted. Often, sensor data require freshness that conflicts with caching. Furthermore, there is also the demand for unscheduled traffic in the IoT e.g.,  the control of actuators, which is much easier to achieve in an end-to-end access model. Finally, in many ICN approaches, routing and forwarding significantly increases the burden over IP. In effect, state and cached content may blow up memory requirements of constrained nodes. 

At the conceptual level, it remains fairly open whether benefits outweigh the shortcomings of ICN in the IoT, or not.
It is the objective of the present paper to explore the basic feasibility and tradeoffs in an experimentally driven approach.    

\subsection{Related Work}

While several ICN approaches have been developed, including NDN \cite{Jacobson:2009:NNC:1658939.1658941}, PSIRP \cite{PSIRP}, Netinf \cite{Dannewitz2013721}, DONA \cite{DONA}, a number of key aspects remain challenges for ICN \cite{draft-icnrg-challenges}. One example of such challenge is the design of routing schemes enabling automatic, efficient, and scalable forwarding information configuration on each ICN device. Related work proposed routing approaches based on proactive, link-state mechanisms \cite{Zhang-Link-State-NDN}, \cite{ospf-ndn}. However, such approaches may not be directly applicable in the IoT, where constrained devices impose different requirements in terms of memory and power capacities. For instance, requirements for home, industrial and building automation \cite{rfc5867} led to the design of the RPL \cite{rfc6550} routing protocol, which can be more energy and memory efficient than standard link-state approaches. It does not require periodic flooding and allows partial topology knowledge.

Recent work has thus started to study ICN paradigms in IoT scenarios or similar contexts (e.g. mobile ad hoc networks). In \cite{CCNx-Contiki}, authors reports on early efforts to provide constrained devices with a CCN communication layer in practice. This implementation is however not interoperable with the full-blown, reference CCN implementation. This initial implementation was used in  \cite{CCN-Homenet-SIGCOMM} to showcase a health monitoring application prototype in the context of a small home network. Several architecture design proposals emerged recently for ICN in the Internet of Things, such as \cite{M2M-IoT-Arch} which proposes an overlay ICN architecture designed over the M2M ETSI standard, or  \cite{ICN-Architecture-IRTF} which identifies high-level requirements of ICN for IoT and proposes a network architecture for IoT based on ICN. Other efforts have proposed enhancements to tackle various issues with ICN in wireless scenarios. For instance, \cite{yu2013interest} focuses on MANETs scenarios and mobile nodes using ICN and proposes a mechanism reducing the overhead of NDN packet forwarding. On the other hand, \cite{amadeo2013named} focuses on sensor networks and data collection from a data sink, and proposes in this context an NDN extension for directed diffusion with new packet types and neighbor distinction. This implementation is however not interoperable with the reference CCN implementation. In \cite{CCN-traffic-optimization} authors propose a push mechanism for CCN, targeting sensor networks. In \cite{Gerla-Gossip-NDN-2012} a gossip mechanism for CCN is introduced, targeting wireless ad hoc networks. Another category of efforts have focused on tackling security and naming issues with ICN in the IoT, such as \cite{burke2012securing} which studies such issues with CCN in the context of lighting systems and building automation.

However, the above prior work only studied ICN approaches via theoretical analysis and simulations. In \cite{CCN-Homenet-SIGCOMM}  and \cite{CCNx-Contiki}, preliminary tests are reported on small, toy networks. But to the best of our knowledge, there are no reports to date on larger scale deployments on IoT hardware, in environments matching requirements described by the industry, e.g., in \cite{rfc5867}. Furthermore, prior work in this domain has either (i) focused on MANET, where machines are not constrained devices, or (ii) focused on wireless sensor networks and sink-centric data traffic (i.e., sensor-to-sink or sink-to-sensor) which is not representative of the whole IoT, where other types of devices participate, and other types of data traffic are significant, such as sensor-to-sensor traffic which is substantial in building automation scenarios (e.g., for lighting systems). 

\subsection{Contributions of this Paper}

In this paper, we report on the first CCN experiments in a life-size IoT deployment, spread over tens of offices on several floors of a building, matching characteristics and requirements  from building automation as specified in \cite{rfc5867}. Based on the insights gained with these experiments, the paper analyses the shortcomings of NDN applied to IoT. Several interoperable CCN enhancements are then proposed and evaluated, which decrease interest traffic and focus data path and caching to match IoT requirements in terms of energy and bandwidth constraints, and  increase content availability in case of IoT nodes with intermittent activity. This paper also provide the first experimental comparison of CCN with the alternative dominant approach in IoT based on 6LoWPAN/RPL/UDP. In addition to our real-world experiments, we discuss ICN in the context of IoT, based on an extensive literature survey.

The remainder of this paper is organized as follows. First, in \S~\ref{sec:ICN-IoT-Issues} we will compare IoT requirements with basic ICN characteristics to identify mismatches and challenges one faces with ICN in the Internet of Things. Then, in \S~\ref{sec:NDN-Deployment} we will describe our ICN implementation for the IoT and our deployment setup in a building automation context. Based on insights gained from our experiments with the CCN implementation in this deployment, we will propose and evaluate in \S~\ref{sec:NDN-Enhancements-Experiments} several interoperable enhancements for CCN operation in the Internet of Things. We will then present lessons learned in \S~\ref{sec:lessonslearned}. Finally, we will conclude and discuss future steps in \S~\ref{sec:conclusion}.

\section{A Priori Challenge of ICN in IoT: Limited Memory}
\label{sec:ICN-IoT-Issues}


Limited memory resources are fundamental in IoT scenarios. Before an ICN solution can be deployed and experimented with, it needs to be aligned with these constraints. In this section, we discuss memory requirements introduced by ICN and how we overcome this basic challenge. We separately discuss aspects concerning caching, protocol stack architecture, and routing schemes. For challenges we derived based on our experiments, we refer to Section~\ref{sec:lessonslearned}.

\subsection{Implications on Caching Capabilities}
One of the fundamental aspects of ICN is in-network caching, which requires memory dedicated to content cache on nodes in the network. On constrained devices, available RAM is very limited and usually in the order of 10~kBytes \cite{rfc7228}. This memory is shared by all processes running on the device, including the operating system, the full network stack, the application(s). Considering typical sizes of these software components in the IoT, the remaining cache size for content on constrained devices is at most in the order of 1~kByte. This is extremely small compared to cache sizes expected on types of devices initially targeted by ICN \cite{ano-crdi-10,pv-rcccn-11}. As readings of sensor values are ephemeral information by nature -- sensor data are continuously replaced by new data -- one might argue to disable caching altogether. However, as we will show below, caching is not only doable, but also beneficial in the IoT (even with such limited resources).

First, a significant part of the data is expected to consist in small size content. The size of a common implementation of temperature values is 12~bytes, which allows to store $\approx$85 sensor values in a single cache. For medium-sized content (i.e., of size in the order of $n$ kBytes, where $n$ is the number of nodes in the network),
distributed caching strategies could coordinate multiple devices to leverage in-network caching of all chunks. Typical medium-size content examples include accumulated, periodically-generated data, or software update binaries.

Second, beyond simple sensor scenarios with a single sink, the IoT envisions multiple consumers for the same content. For example, a temperature sensor asynchronously accessed by the air-conditioning system, the automated blinds, and windows of a room, each of which may react independently upon temperature evolution.  For more powerful devices crowd computing \cite{mych-ccc-10} is an interesting application field. Similarily, caching ephemeral content within the network may significantly increase content availability because (i) nodes can then sleep as often as possible to save energy, and (ii) lossy multi-hop wireless paths towards content producers are shortened. We will study the effect of caching in Section~\ref{sec:NDN-Enhancements-Experiments}.


\subsection{Implications on Overlay Applicability}
\label{sec:challenges-overlay}
Deploying only the IP stack on constrained devices is already a challenge in terms of RAM and ROM. ICN approaches that work on top of IP might be impossible due to the additive memory requirements of both the ICN stack and the IP stack. Consequently, ICN implementations should work directly on top of the link layer. Note that for heterogeneous deployment border gateways can bridge between IP and ICN. For the experiments reported in this paper, we have thus used an ICN approach running directly above the MAC layer (see Sections \ref{sec:NDN-Deployment} and \ref{sec:NDN-Enhancements-Experiments}). 

\subsection{Implications on Routing Approaches}
Reduced memory of constrained devices also limits applicability of ICN routing approaches. Current proposals usually route either directly on names or indirectly via name resolution. 
Based on our previous observations, name resolution on top of IP is not viable.
However, even some pure name-based routing schemes, such as \cite{Zhang-Link-State-NDN} and \cite{ospf-ndn} rely on an ICN overlay requiring an IP network, or use proactive link state algorithms. Link state routing results in both (i) a significant amount of control traffic, whether or not there is data traffic to carry in the network, and (ii) a significant amount of memory, typically in O($n$), where $n$ is the number of nodes in the network. These characteristics do not match the memory and energy resources of constrained devices. 

Routing protocols running on IoT devices should aim for O(1) routing state and minimal control traffic -- ideally none, especially when there is no data traffic to carry \cite{levis2009overview}. In Section~\ref{sec:NDN-Enhancements-Experiments}, we introduce ICN routing with these~properties.

\section{Steps to Enable ICN in the IoT}
\label{sec:NDN-Deployment}

In order to gain a full understanding of how ICN operates in the Internet of Things, it is inevitable to conduct experiments in real-world deployments or testbeds that reflect properties of such deployments. Testbeds help to avoid topologies and densities that are too artificial, too regular, or too isolated compared with the real word. They naturally include external interferences resulting from other radio networks, electrical devices, or simple human activity. The first step towards such experiments is implementing ICN code that runs on IoT hardware. 

\begin{figure*}
\centering
\includegraphics[width=\textwidth]{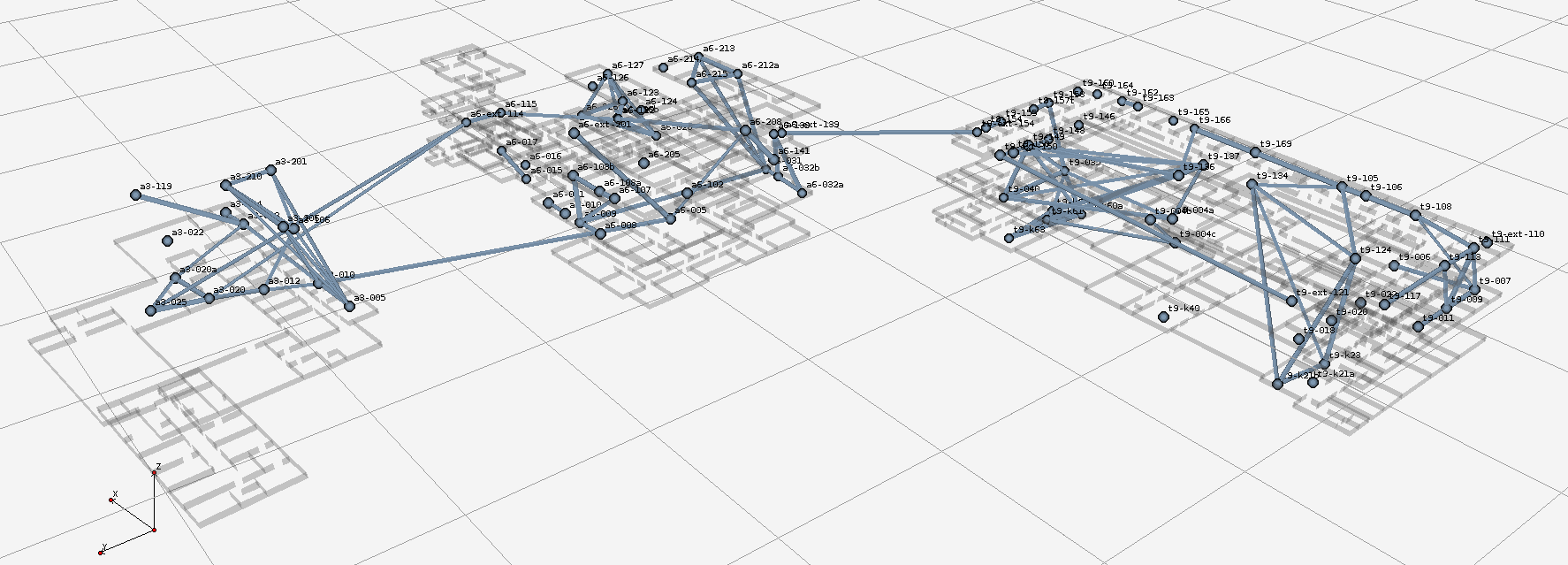}%
\caption{3D visualization of the topology of the deployment, consisting in 60 nodes that interconnect via wireless communications (sub-GHz) and that are physically distributed in multiple rooms, multiple floors, and multiple buildings.}%
\label{fig:topology}%
\end{figure*}

\subsection{Porting CCN-Lite to RIOT}

We have ported CCN-Lite \cite{ccn-lite}, a bare-bone Linux open source implementation of NDN, to RIOT  \cite{RIOT-INFOCOM}, an operating system for constrained devices. Among ICN approaches, we have chosen NDN because it can easily operate directly above the link layer -- a requirement we identified in Section~\ref{sec:ICN-IoT-Issues}. We chose to base ourselves on CCN-Lite because this implementation is compliant with the reference NDN implementation (CCNx) while being very compact: less than 1,000 lines of C code and low memory footprint. And we chose RIOT as operating system to run on constrained devices because it is open source and fits IoT devices memory requirements, while allowing plain C code with all the standard headers, based on a (multi-)threading model comparable to POSIX. These characteristics guaranteed that porting Linux code to RIOT is straightforward, and a fair comparison with the non-IoT world. We also leveraged RIOT support for popular debugging tools such as Valgrind, Wireshark, gdb, and nativenet. Our implementation is open source and available online in GitHub \cite{RIOT-github}. 

Table~\ref{tbl:memory-resources} compares the ROM and RAM sizes of the binaries compiled for NDN network stacks and for 6LoWPAN/RPL network stacks, built upon state-of-the-art IoT operating systems (RIOT and Contiki), for state-of-the-art IoT hardware (Redbee Econotag board and MSB-A2 board). We observe that an ICN approach can significantly outperform common IoT protocols in terms of ROM size (down to 60\% less) and RAM size (down to 80\% less).

\begin{table}[t]
\centering
\subtable[RIOT on MSBA2]{
\begin{tabular}{ l c c }
    \toprule
    \textbf{Module}  & \textbf{ROM}  & \textbf{RAM}  \\
    \midrule
    RPL + 6LoWPAN \quad  & 53412 bytes \quad   & 27739 bytes \quad \\
    CCN-Lite         & 16628 bytes   & 5112 bytes      \\
\bottomrule
\end{tabular}
\label{tbl:resources-riot}
}
\smallskip
\subtable[Contiki on Redbee-Econotag]{
\begin{tabular}{ l c c }
\toprule
    \textbf{Module}          & \textbf{ROM}       & \textbf{RAM}       \\
    \midrule
    RPL + 6LoWPAN \quad  & 52131 bytes \quad    & 21057 bytes \quad   \\
    CCNx            & 13005 bytes     & 5769 bytes     \\
\bottomrule
\end{tabular}
\label{tbl:resources-contiki}
}
\caption{Comparing memory resources for common IoT operating systems and hardware.}
\label{tbl:memory-resources}
\end{table}

\subsection{Configuring NDN Deployment}
\label{sec:ndn-deployment}

In order to obtain a fully functional NDN network stack for the IoT, a FIB autoconfiguration mechanism is needed: in IoT scenarios, even less than in other scenarios, one cannot expect humans in the loop, so manual configuration is not part of the deployment. In particular, predefined location-based naming and simple routing schemes based on the structure of such names may thus not be possible in general. Furthermore, as mentioned in Section~\ref{sec:ICN-IoT-Issues}, existing ICN routing approaches are not appropriate for constrained devices in the IoT: alternative routing mechanisms must be used in this context, which require drastically less state.

In the context of ICN, the naming scheme is crucial. NDN uses a hierarchical name space, which allows for aggregation in routing. The amount of content items that can be expressed depends on the character set and name length. MTUs of common IoT link layer technologies range between $\approx$30~bytes and $\approx$100~bytes. To the best of our knowledge fragmentation within ICN is not addressed, hence naming and chunk size need to be aligned with the packet size to prevent fragmentation (not supported by the link layer).

\section{NDN Experiments and Optimizations for IoT Deployment}
\label{sec:NDN-Enhancements-Experiments}

In the following, we will describe and evaluate several routing alternatives, as well as other aspects of NDN in the wild, such as the effect of caching in IoT.

\subsection{Large-scale Deployment Setup}
\label{sec:deployment-setup}
Typical IoT application scenarios, include building and home automation~\cite{rfc5826, rfc5867}, smart metering (e.g., \cite{fan:2010}), or environment monitoring (e.g., \cite{wittenburg07fence}). These scenarios usually require a multi-hop wireless network. For the NDN experiments, we deployed our ICN IoT implementation on the campus testbed of \emph{Freie Universit\"at Berlin}, consisting in 60 nodes distributed in various rooms, on several floors, and in several buildings, as shown in Figure~\ref{fig:topology}. This deployment matches the typical device density (several meters between nodes), distribution (one node per room), and environment (e.g., co-located wireless networks) described in \cite{rfc5867} for building automation, e.g., HVAC devices, lighting devices, or fire-detection devices. Each node is equipped with a CC1100 radio chip operating at 868MHz, and sensors that can measure various parameters including room temperature, humidity etc. For more details we refer to~\cite{Baar+:2008}. Most of the nodes are deployed inside rooms, while a few nodes are deployed outdoor to better interconnect nodes in different buildings. Nodes interconnect via their wireless interface, which offers a maximum link layer frame size of 64~Bytes.

In order to monitor closely energy consumption, verify individual node behavior, and manage experiments on this deployment (e.g., flash nodes, gather results) each node is furthermore connected to its own docking station. Docking stations are interconnected via an Ethernet backbone. However, these docking stations are used only to monitor and manage the nodes. Nodes operate autonomously, i.e., each node can only use its own CPU, its own memory, and its own wireless interface to communicate with other nodes.

\begin{figure*}[t]%
\centering
\subfigure[10 nodes are involved when a single consumer (t9-k38) requests content published by t9-155.]{\includegraphics[width=0.90\columnwidth]{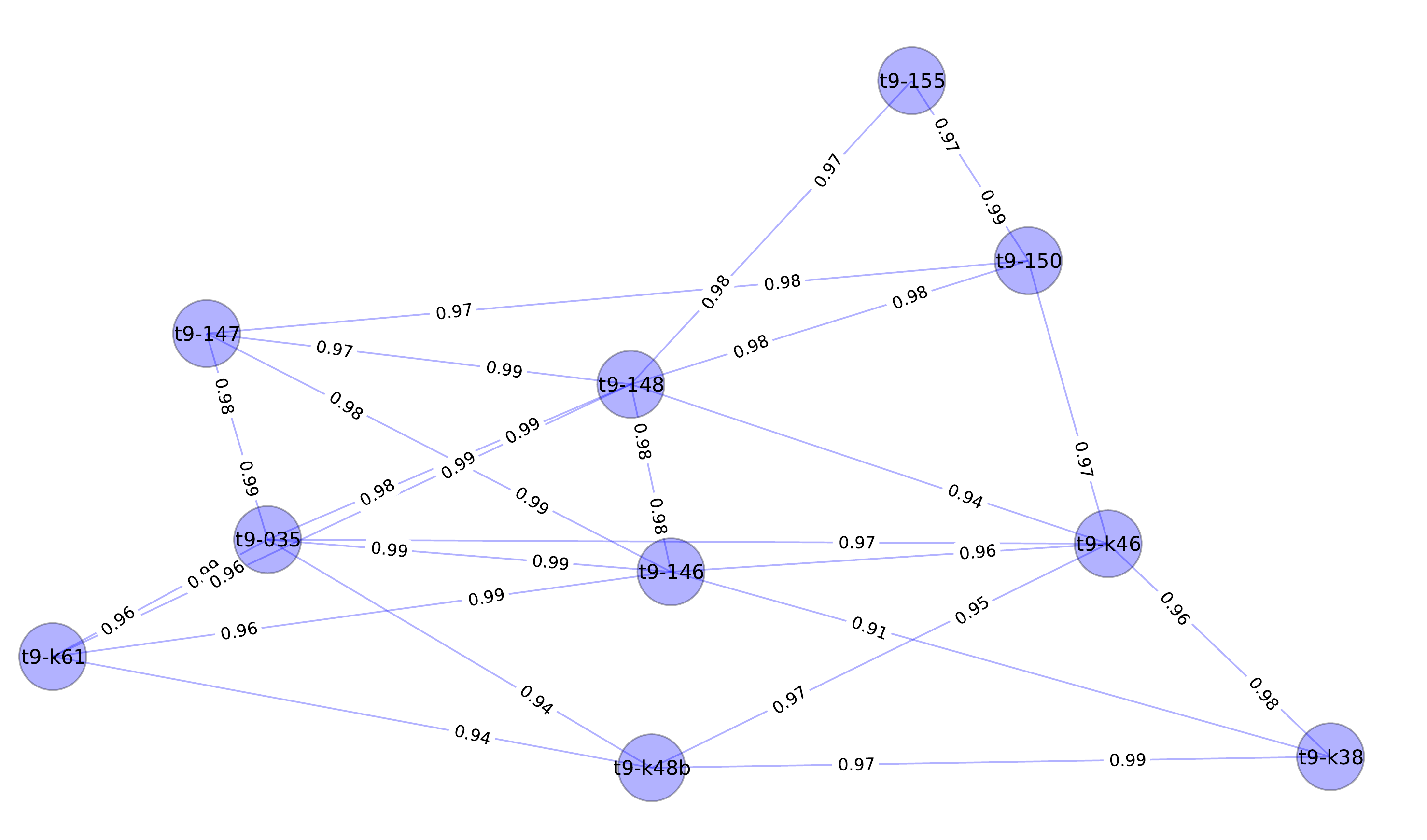}\label{fig:10-nodes-topo}}
\qquad
\subfigure[20 nodes are involved when multiple consumers (t9-149, t9-148, and
t9-150) request content published by t9-k36a]{\includegraphics[width=0.90\columnwidth]{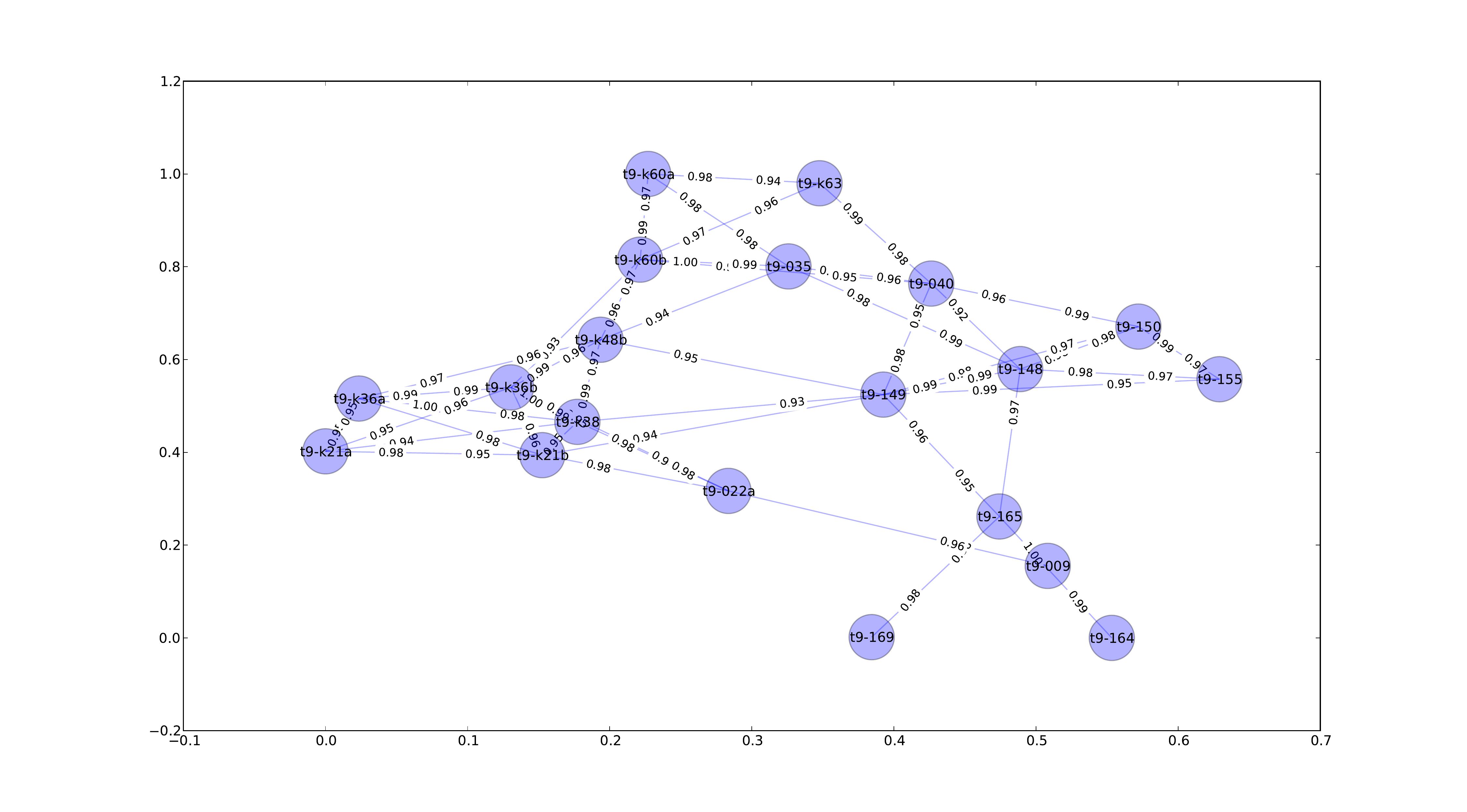}\label{fig:20-nodes-topo}}
\caption{Snapshot of the link-layer network topologies used in the experiments for single and multi consumer scenarios. Each topology spans over 3 floors in the right-most building shown in Figure~\ref{fig:topology}. Link weights describe \% of received packets, per link, per direction.}%
\label{fig:tested-topologies}%
\end{figure*}

\smallskip\noindent
\textbf{Basic Configuration of Experiments}\quad
The following experiments use 400~ms interest timeout (stop-and-go, giving up after 5~tries), and 900~ms nonce timeouts. The content is named in a hierarchical fashion typical for NDN, without any encryption. Considering the maximum link layer frame size of 64~bytes in our deployment, we decide for a medium sized name length of 12~bytes including the chunk identifier (the exact names of the content chunks are \emph{/riot/text/a}, \emph{/riot/text/b} etc.). Note that with these names, the size of headers and names fit in a single link layer frame, both with CCN (16+12 = 28~bytes) and with 6LoWPAN/RPL/UDP (15+12 = 27~bytes), and still allow to carry realistic application data. Also note that the sizes of minimal CCN header (16 bytes, eliding optional fields) and of 6LoWPAN/RPL/UDP headers (15 bytes) are similar, and thus represent not a decisive factor in the differences observed in the following experiments. The length of content names is however a factor, as discussed in Section~\ref{sec:energy}. 

In the experiments, we consider a single content producer and one or multiple consumers. Due to the volatile nature of the wireless medium~\cite{baccelli2014multihop}, the resulting link layer topologies based on our 60~node network might change on a per-transmission basis (cf., Figure~\ref{fig:tested-topologies}). Note that IoT scenarios in home and building automation networks are typically multi-hop, but less than 5 hops in diameter \cite{goyal2013reactive}. Consequently, in our experiments, we placed content producer and consumers at least 2 hops apart.

To analyze the effects of NDN for typical radio packets payload in the IoT, we align the chunk size such that each chunk can be transmitted without fragmentation. In our case, MTU is 64 bytes, chunks are set to be 58 bytes long, of which 30 bytes of content. Since typical sensor content production is of the order of 200~bytes per minute \cite{rfc5867}, we set the basic configuration for consumers to periodically fetch 10 such chunks. However, other popular IoT radio technologies provide MTUs that are twice bigger (e.g. IEEE 802.15.4), or half smaller (e.g. Bluetooth LE). So we also check cases with 5 and 20 chunks per content item.




\subsection{Vanilla Interest Flooding (VIF)}

The simplest routing approach that requires minimal states is \emph{interest flooding}, whereby each node in the network repeats an interest, upon first reception. In the following, we will call this simple mechanism Vanilla Interest Flooding (VIF). Using VIF, a consumer with an empty FIB can nevertheless disseminate its interest in content, and the flooded interest will reach the producer which can then send the content on the reverse path. VIF fits the constraints of IoT devices because (i) it does not rely on any additional control traffic to maintain the FIB, (ii) it requires minimal state, i.e., only temporary pending interests on the reverse path of content that is sought after.

\begin{figure*}[t]%
\centering
\subfigure[Vanilla Interest Flooding]{\includegraphics[width=.95\columnwidth]{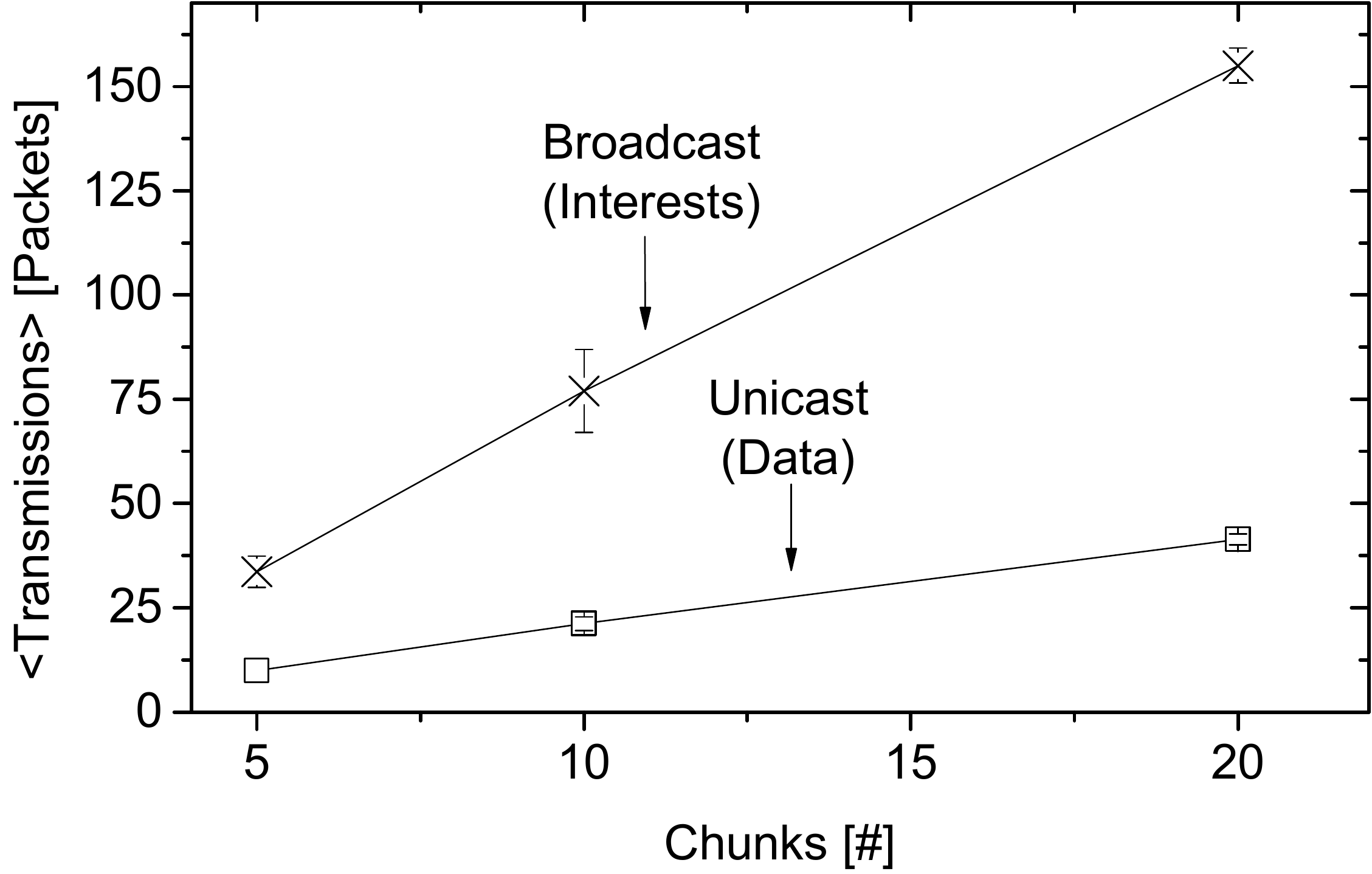}\label{fig:no-routing}}
\hfill
\subfigure[Reactive Optimistic Name-based Routing]{\includegraphics[width=.95\columnwidth]{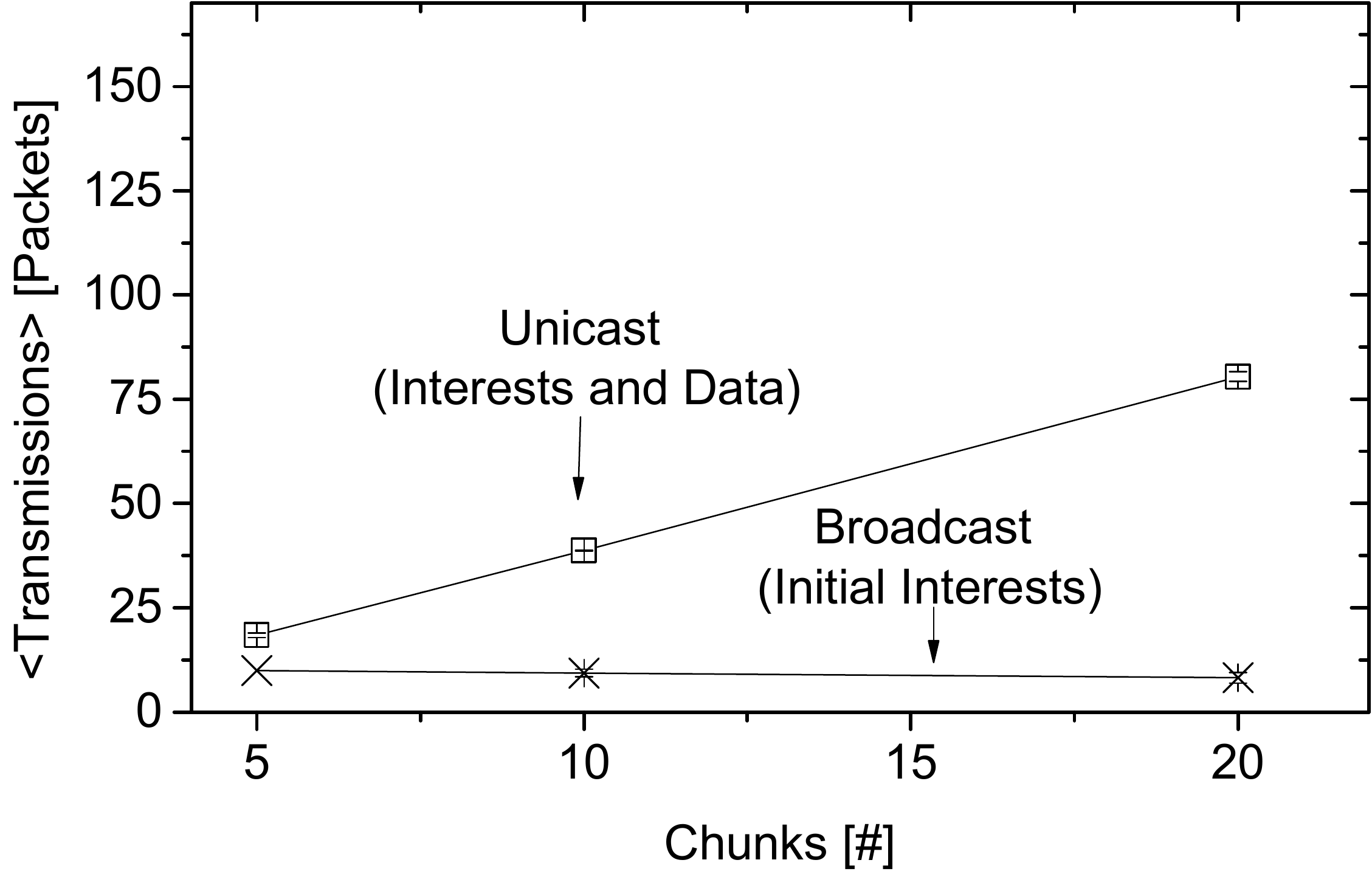}\label{fig:routing}}
\caption{Single-consumer scenario. NDN performance for different routing schemes. Average number of packets transmitted in a network of 10 nodes to fetch content of various size.}%
\label{fig:routing-comparison}%
\end{figure*}

Figure~\ref{fig:no-routing} shows the results of an experiment using NDN with VIF for a single consumer scenario. In this experiment, the consumer periodically accesses content of size 5, 10, or 20 chunks of data, all of which were produced by another constrained node in the network shown in Figure \ref{fig:10-nodes-topo}.



While the experiment is successful in that NDN was demonstrated to operate on IoT hardware (meeting memory requirements), and the consumer could fetch the content, Figure \ref{fig:no-routing} shows that, compared to its size, many packets were transmitted to fetch the content. This is due to the fact that each chunk triggers an interest, which requires network-wide flooding.
In general, in a network of $n$ nodes, and for $k$ chunks of content, the number of transmissions for a single content item is $k\cdot((n-1)+\sqrt{n})$, assuming the average path length approximation $\sqrt{n}$. We observe that while VIF is simple and works in the scenario we tested, it does not scale well in terms of number radio transmissions when the network or the content grows in size. Radio transmission and reception are however very costly in terms of energy for battery-powered IoT devices. In the following, we have thus designed and tested enhancements reducing the number of radio transmissions and receptions in IoT environment.



\begin{figure*}[t]%
\centering
\subfigure[Without caching]{\includegraphics[width=.95\columnwidth]{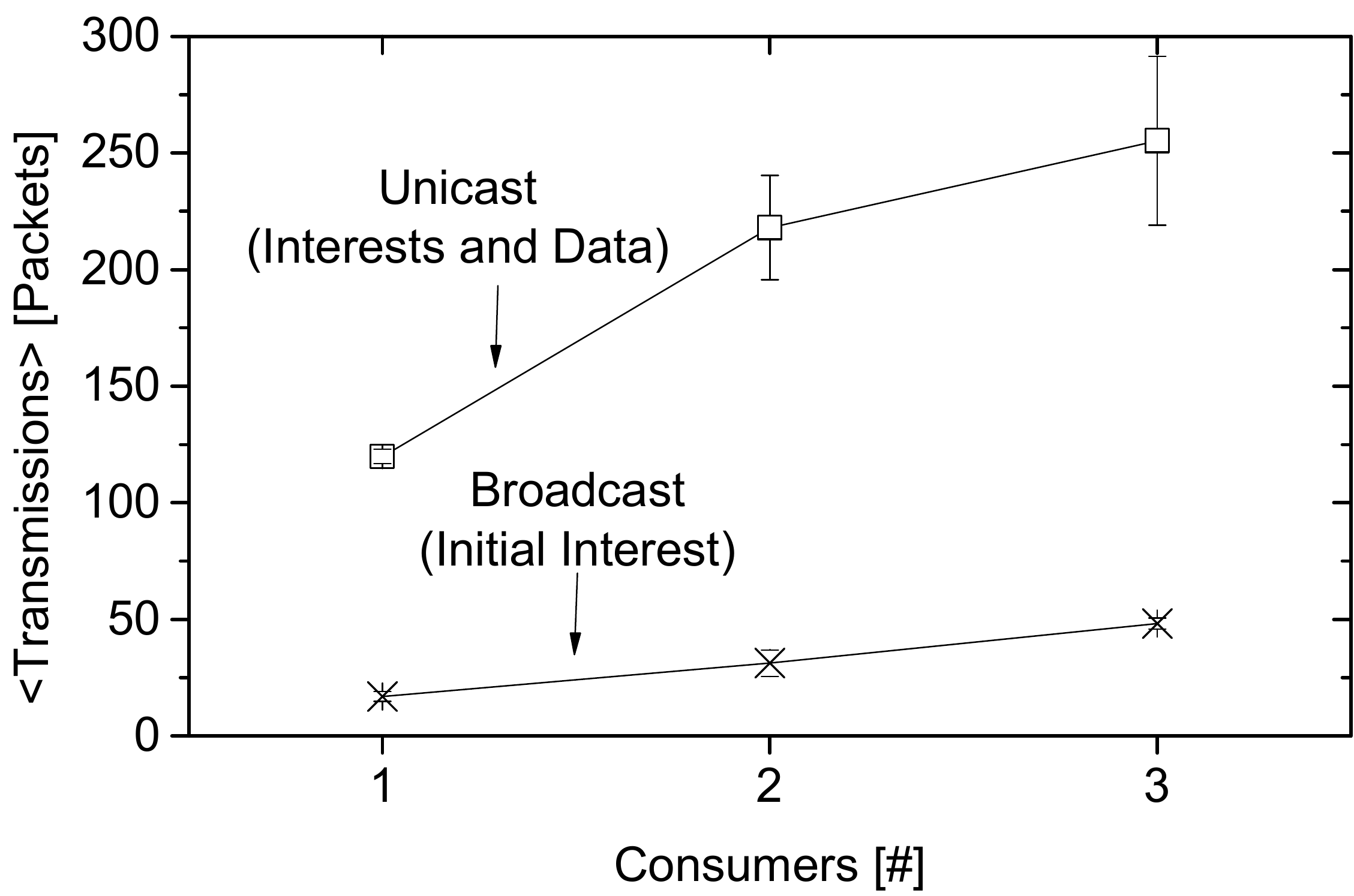}\label{fig:no-cache}}
\hfill
\subfigure[With caching]{\includegraphics[width=.95\columnwidth]{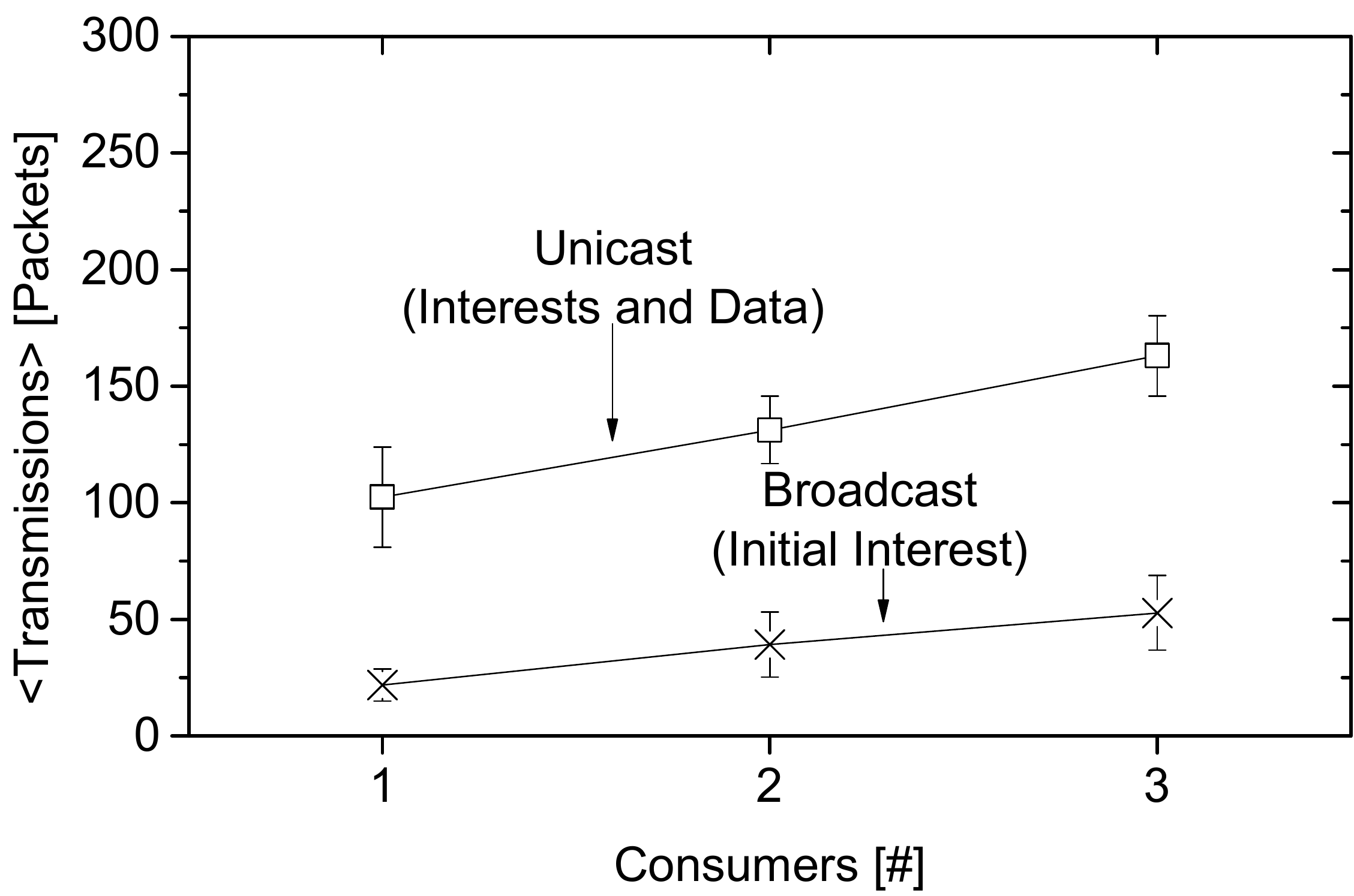}\label{fig:cache}}
\caption{Multi-consumer scenario. NDN performance for RONR and different content cache schemes. Average number of packets transmitted in a network of 20 nodes with a variable number of consumers.}%
\label{fig:cache-comparison}%
\end{figure*}

\subsection{Reactive Optimistic Name-based Routing (RONR)}

In order to reduce the number of radio transmissions compared to basic interest flooding, we introduce Reactive Optimistic Name-based Routing (RONR), which automatically configures a temporary FIB entry on the reverse path taken by the first content chunk. That way, in case the FIB is empty (e.g., after booting) or if no FIB entry matches the name/prefix of the content in which the consumer is interested, only a single initial interest flooding is needed, while subsequent interests for chunks of that content can be unicast using the FIB entries thus auto-configured along the path. For example, in our experiments, after flooding an interest for chunk \emph{/riot/text/a}, nodes on the reverse path of that chunk store a temporary FIB entry for \emph{/riot/text/*}, thus subsequent interests for chunks \emph{/riot/text/b}, \emph{/riot/text/c} can be unicast using the established path, instead of flooded. 


RONR is optimistic because it first assumes that the whole content is stored on a single node (a cached replica or the original producer), which may not be the case in general. However, this assumption is reasonable in the IoT because typical content size is in the order of a few hundred bytes \cite{rfc5867}. Furthermore, FIB entries timeout ensure that if the configured FIB entries do not lead to a node with the full content, the consumer will eventually revert to interest flooding, through which it can discover another node with the rest of the content, install new temporary FIB entries etc. This timeout strategy is common for reactive routing in multi-hop wireless scenarios \cite{richard2005defining}.

In Figure~\ref{fig:routing}, we show the results of an experiment using NDN with RONR, for the exact same topology and scenario as for Figure~\ref{fig:no-routing}. We observe that the number of radio transmissions decrease about 50\% compared to NDN with VIF. In particular the number of broadcast transmissions is drastically reduced because, with RONR, only the first interest packet of a content item is flooded, while subsequent interests are unicast, using temporary FIB entries established by RONR. A quick back-of-the-envelope analysis shows that in a network of $n$ nodes, and for $k$ chunks of content, the number of transmissions is $(n-1)+2(k-\frac{1}{2})\sqrt{n} $, assuming again the average path length approximation $\sqrt{n}$. Therefore, RONR scales much better than VIF when network size or content size grows. RONR thus better fits IoT devices energy requirements compared to VIF, while still fitting other requirements of constrained devices by (i) not relying on any control traffic, and (ii) requiring minimal state, i.e., only temporary FIB entries on the reverse path of content that is sought after (not counting PIT state, of course). 

An enhancement of RONR could be even more optimistic and tentatively aggregate prefixes in the following manner. If a FIB entry is pointing to an interface for content with prefix \emph{/riot/text/*} and an interest for \emph{/riot/temp/c} is answered by a chunk of content through the same interface (after the initial interest flooding phase), the enhancement would optimistically aggregate the prefixes and create a FIB entry for \emph{/riot/*} pointing to this interface. In the best case, this will indeed lead to all the requested content matching this prefix, via unicast only. In the worst case, after unicast transmission and time-out, the consumer will eventually revert to interest flooding, through which it can discover another node with the rest of the content, install new temporary FIB entries etc. For this paper, however, we have only tested RONR without this enhancement, and leave its analysis for future work.

\subsection{Multiple Consumers \& Impact of Caching}
\label{sec:caching}

In this section, we evaluate experimentally the impact of ICN caching. The same content (20 chunks) is accessed alternatively by one, two, or three consumers that are topologically close to one another (pairwise, maximum hop distance is 1). In order to accommodate for more consumers while keeping them apart from the producer with at least 2 hops, a larger topology shown in Figure \ref{fig:20-nodes-topo} was used for the following experiments. To reduce signaling overhead, we use RONR as routing scheme for NDN interest packets.

In Figure~\ref{fig:no-cache} we show the results of our experiment with a disabled content cache.  We observe that, as expected, the number of radio transmissions scales almost linearly with the number of consumers. In a network of $n$ nodes, and for $k$ chunks of content and $m$ consumers within radio reach, the number of transmissions is $m\cdot((n-1)+2(k-\frac{1}{2})\sqrt{n}) $, still assuming the average path length approximation $\sqrt{n}$.

Next, we enable cache capacity of 20~chunks on all nodes, which corresponds to RAM usage of 2~kBytes (2~\% of 96~kbytes overall RAM). Figure~\ref{fig:cache} shows the results we obtained for the exact same topology and scenario as for Figure~\ref{fig:no-cache}, except the caching. We observe that the number of radio transmissions needed to retrieve the content is drastically reduced, by up to 50\% in this scenario. In detail, the number of broadcast transmissions is almost similar, while the number of unicast packets decreases substantially. This is consistent with the facts that the initial interest flooding (broadcasted) is not modified, while cached content chunk shorten unicast paths, thus reducing the number of unicast transmissions. In the best case, if the initial flood for subsequent consumers can be reduced to a local broadcast because only neighbors with cached content receive the interest, the number of transmissions becomes $2(k-\frac{1}{2})(\sqrt{n}+n-1)+n+m-2 $. 

\subsection{Comparison with 6LoWPAN/RPL/UDP}

In this section, we compare NDN with 6LoWPAN/RPL/UDP, a common protocol suite for the current IoT. For fair comparison, we use the following setup. On the ICN side, we deploy RONR with a cache size of 2~kBytes, as this leads to the best performance results in our previous analysis. On the RPL side, we first let the network converge until the RPL root and the routing entries are installed in nodes, before we start the experiment (i.e., we factor out the control traffic transmissions necessary to bootstrap the network).

\begin{figure}[t]%
\centering
\includegraphics[width=1.0\linewidth]{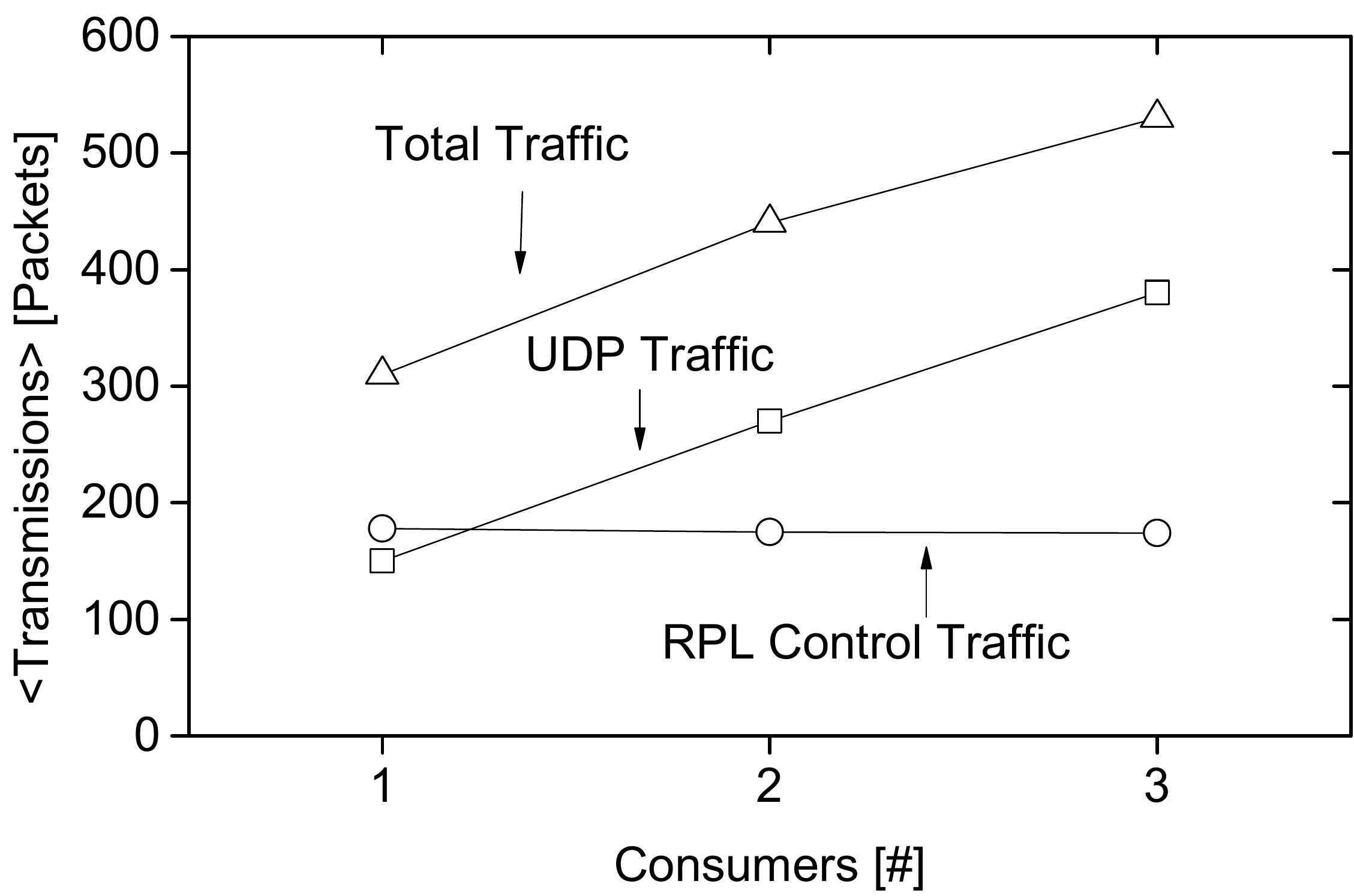}%
\caption{Multi-consumer scenario with 6LoWPAN/RPL/UDP. Average number of packets transmitted in a network of 20 nodes.}%
\label{fig:rpl}%
\end{figure}

In Figure~\ref{fig:rpl}, we show the results we obtained for the exact same topology and scenario as for Figure \ref{fig:cache}, except the network stack used was 6LoWPAN/RPL/UDP with default settings instead of NDN. We observe that the 6LoWPAN/RPL/UDP network stack yields much more transmissions compared to NDN (cf., Figure \ref{fig:cache}), approximately three times more. This is due to two main factors. On one hand, the amount of control traffic generated by the \emph{proactive} 6LoWPAN/RPL/UDP network stack is a big penalty compared to the \emph{reactive} CCN approach we tested. On the other hand, compared to our CCN approach, the unicast paths created by the 6LoWPAN/RPL/UDP network stack do not benefit from \emph{caching} and are thus always maximum length, which can in some cases be even longer than the shortest topological paths, as shown in \cite{xie2010performance}. Note that we have not used RPL extensions such as \cite{goyal2013reactive}, which could reduce the length of unicast paths. Furthermore, as discussed in Section~\ref{sec:deployment-setup}, we observed that the naming scheme and the header sizes were not a decisive factor explaining the performance gap between the CCN stack and the 6LoWPAN/RPL/UDP in the experiments we conducted.  All in all, we can conclude that NDN may be a potential alternative to 6LoWPAN/RPL/UDP, which should be studied more in the context of IoT in future work.


\section{A Posteriori Challenges: What are the Lessons Learned}
\label{sec:lessonslearned}

In this section, we gathered further considerations and observations concerning ICN in the Internet of Things, based on our practical experience with NDN implementation and deployment. In the following, we distinguish energy consumption aspects, wireless connectivity aspects, and communication model aspects.

\subsection{Energy Consumption}
\label{sec:energy}
Energy consumption is mainly impacted by network transmissions, which are affected by content naming, content caching, network flooding, and local wireless broadcast.

\subsubsection{Impact of Names}
Routing information about names and prefixes should dynamically be auto-configured in IoT devices. The resulting overhead not only depends on the routing protocol but also on the size of names to be processed in ICN packets. In our experiments, we deployed VIF, a very basic approach based on \emph{flooding}, whereby each node in the network repeats (on all interfaces) each flooded packet upon first reception (on any interface). 

Flooding is used (i) to disseminate an interest message when no forwarding information is available, or (ii) to disseminate names and topology information, e.g., with link state routing approaches \cite{Zhang-Link-State-NDN}, \cite{ospf-ndn}. However, flooding is costly in terms of energy since each flood requires O($n$) packet transmissions and O($nm$) packet receptions, where $n$ is the number of nodes in the network and $m$ is the average node degree. Each packet received will not only be costly in terms of pure packet reception but will also trigger its processing, which includes CPU-expensive string comparisons with variable lengths, trying to match received names with names stored locally. Furthermore, recent work \cite{wsv-bdpts-13} identifies ICN packet processing as a CPU bottleneck, serious enough to provide DOS attack opportunities. This processing is even more costly on constrained devices since their CPU typically does not benefit from advanced functionalities such as prefetching or super scalar instruction set, and thus needs one cycle per byte compared. Table~\ref{tbl:ccn-cpu} shows a benchmark for the number of required CPU cycles per CCNlite operation for our implementation in RIOT. The top 3 functions, which represent 85\% of the CPU cycles, involve string comparison and name matching. 

These observations thus call for (i) the least possible recourse to flooding and (ii) the shortest possible names. Note that short names also ensure that packet fragmentation is avoided at the link layer: long names that do not fit in the MTU of the link layer split interest packets in several transmissions which is inefficient in terms of energy. Shorter names should however not sacrifice prefix aggregability, so that scalability remains in terms of number of nodes in the network vs. routing state. In our experiments, we have demonstrated the use of small, hierarchical names of 12~bytes and the minimal CCN header all of which carried by link layer packets with very constrained MTU (64 bytes). In practice, it still allowed about 30 bytes of content payload, which is appropriate for IoT scenarios, where content generated by sensors is in the order of a few hundred bytes per minute \cite{rfc5867}. Even with slightly longer names implementing a typically deeper hierarchy as described in \cite{rfc5867}, there is still enough space for payload. For instance, with names such as \emph{/zone1/room2/dev7/temp/a}, there is about 20 bytes for content payload with the link layer we used in our experiments.

Note however that human-readable names may not be required or useful in a context of machine-to-machine communication, whereby no humans are in the loop. More compact naming (e.g., a binary representation, or a more compact ASCII representation) may thus be applicable and would leave more space for content in packets constrained by small MTUs. Furthermore, the computations incurred by cryptographically-generated names (or parts of names) are expected to yield both substantial energy consumption penalty for constrained node in the IoT. In this paper, however, we do not consider security-related aspects for names -- which are considerations that are orthogonal to the aspects studied here. Nevertheless, security mechanisms typically yield substantially longer headers. We can therefore also conclude that a standard CCN header compression scheme would be useful in the IoT.



\begin{table}
\centering
\begin{tabular}{r l}
\toprule
\textbf{\# of instructions}      & \textbf{Function}  \\
\midrule
14,002,814                  & memcmp\_ssse3 \\
 7,525,050                  & ccnl\_nonce\_find\_or\_append \\
 4,062,659                  & ccnl\_i\_prefixof\_c \\
 1,462,304                  & dehead \\
   956,238                  & ccnl\_core\_RX\_i\_or\_c \\
   895,590                  & ccnl\_extract\_prefix\_nonce\_ppkd \\
   845,042                  & memcpy\_ssse3\\
\bottomrule
\end{tabular}
\caption{CPU cycles per CCN function.}
\label{tbl:ccn-cpu}
\end{table}

\subsubsection{Impact of Caching}
The impact of in-network caching on energy aspects with ICN approaches has been studied by recent work such as \cite{choi2012network}, which indicates that energy consumption incurred by caching reduces energy efficiency. But on the other hand, studies such as \cite{lee2010greening} show that  CCN can be more energy efficient than other content delivery approaches such as CDN and P2P by leveraging the most energy efficient devices in the network. It remains to be seen at large scale on the Internet which ICN approaches introduce low overhead in terms of energy consumption. In the IoT, to the best of our knowledge, there are no studies yet that focused on energy aspects of ICN due to the use of caching.

In Section~\ref{sec:NDN-Enhancements-Experiments}, we demonstrated experimentally that savings in terms of energy consumption are possible thanks to (even small) in-network caching since (i) on-path or near-path caching can decrease the number of intermediate energy-challenged devices on the path to reach content in some scenarios, and (ii) content producers such as sensors could sleep more while their content could still be available in other caches in the network.

\subsubsection{Impact of Local Wireless Broadcast}

In case of multiple PIT hits, the NDN stack could use a single multicast transmission if all matching neighbors are reachable through the same wireless interface -- which is the case in most IoT scenarios where nodes only have a single interface (omnidirectional radio). We have thus enhanced our NDN implementation with such a link-local multicast awareness mechanism called Content Forwarding Aggregation (CFA). In scenarios where multiple geographically close consumers are interested in the same content at approximately the same time, CFA leads to substantial gains in terms of number of radio transmissions necessary to deliver the content. With CFA, a content chunk may be forwarded as a single multicast to multiple nodes that have expressed interest in this content. Using link-local multicast, CFA reaches nodes within the same radio range without implementing explicitly location-awareness mechanisms.


Another opportunity to leverage the multicast nature of IoT devices' wireless interface concerns caching. Very often, a node will overhear unsolicited chunks of content that are being transmitted in its radio vicinity. In such case, instead of discarding this content, the node could cache this unsolicited chunk in its content store, if there is space left, with a lower priority than solicited content. We have thus enhanced our NDN implementation with such a mechanism, called Opportunistic Near-Path Caching (ONPC), which increases availability of the content and further reduces the number of radio transmissions in case of several consumers of the same content. However, due to lack of space, we do not show experimental results with CFA or ONPC in this paper.

\subsection{Wireless Connectivity}

Although ICN is applicable in wireless networks, several issues arise when applied to the wireless regime at work in IoT. In the following, we distinguish aspects concerning frame size, fragmentation, and bidirectional links.

\subsubsection{Frame Size and Packet Fragmentation}
Several link layer technologies are currently used in the IoT, and it is likely that multiple technologies will be used in the future, too. Currently, the dominant IoT link layer in the field of building automation and industrial automation is IEEE~802.15.4. The maximum frame size is very small (127~bytes or less). Other popular wireless link layers provide an even smaller maximum frame size, such as Bluetooth Low Energy \cite{isomaki2013transmission} which typically allows a payload of $\approx$30~bytes. These frame sizes are more than ten to a hundred times smaller compared to traditional Ethernet or WiFi frames. Consequently, fragmentation and reassembly mechanisms are necessary. While Bluetooth provides its own, IEEE~802.15.4 does not. To bridge this gap, 6LoWPAN introduced (i) a standard header compression scheme, and a (ii) standard fragmentation and reassembly mechanism for IPv6 operation in the IoT, both on top of IEEE 802.15.4 link layer.
It is worth noting that ICN cannot benefit from the same mechanisms for fragmentation and compression.
Overlay architectures conflict with the memory constraints in the IoT (cf., Section \ref{sec:challenges-overlay}) as well as with packet sizes of common IoT link layers.

In our real-world deployment, we demonstrated that NDN can be implemented directly on top of an IoT link layer, without compression/fragmentation mechanisms (see Sections~\ref{sec:NDN-Deployment} and \ref{sec:NDN-Enhancements-Experiments}). Omitting these optimizations is suitable for basic scenarios in which small enough names and small enough chunks can be used in the first place. Our results give confidence that we can already start with ICN in the IoT.
However, in the future, ICN approaches for the IoT need an equivalent of what 6LoWPAN is providing for IPv6. 
For illustration, NDN will typically use up to 40 bytes for the header and data encoding, which is negligible in the common Internet ($\approx$2\% of the capacity of standard 1500~bytes MTU) but occupies $\approx$28\% of the capacity of standard 802.15.4 frames.
Neither can it be expected that all chunk sizes on all ICN networks will be defined by IEEE 802.15.4 frame size (which would be inefficient), nor can it be expected that names indicated in interest packets will always be short enough to fit in a single 802.15.4 frame of 127 bytes, for example.
Note that fragmentation approaches need to take into account that altered chunks can break security and naming schemes. 

\subsubsection{Bidirectional links}
Many ICN approaches assume bidirectional links. This is not true in general in spontaneous wireless networks \cite{cordero2013-SIGCOMM-eBOOK}, and thus this assumption does not hold in the IoT. In such context, a high proportion of links are asymmetric, e.g., 10\% loss rate from $A$ to $B$ and 80\% loss rate from $B$ to $A$. In reality, a substantial fraction of the links are unidirectional, i.e., loss rate strictly below 100\% in one direction, and 100\% loss rate in the reverse direction. Last but not least, wireless link quality between two nodes $A$ and $B$ can vary significantly over time, even at small time scales~\cite{baccelli2014multihop} -- a phenomenon we also experienced in our experiments.

The above wireless connectivity characteristics lead to the following observations. ICN routing protocols running on constrained devices need to satisfy conflicting requirements (i) negligible control traffic to reduce energy consumption and small state to fit memory constraints, while at the same time (ii) dynamic tracking of wireless link to avoid non-functional paths. The goal is to not forward an interest in the first place if reverse link is not ``good enough''. The overhead for failing is a reverse path taken by content which often fails and will lead to PIT time-outs, interest flooding, etc. Subsequently, this might lead to the same failing reverse path -- and thus be very inefficient both in terms of energy and delay. 

\subsection{Different Communication Models}
The ICN communication model is based on a \emph{pull} paradigm: in a first phase, a node expresses interest in some content, and in a second phase, the node should receive this content. However, this communication model alone is not sufficient to accommodate typical traffic patterns in the IoT. Aside of pull, these patterns include for instance \emph{push} paradigms (e.g., for actuators), and \emph{observe} paradigms \cite{shelby2014-coap} whereby a node can register for updates from a given content producer (e.g., a sensor measuring the real-time evolution of a given parameter). Note that explicit acknowledgements are also typically used in this context, for example patterns such as push+ACK, or request+reply+ACK are the norm in this domain. Recent work has started to integrate these patterns in ICN, such as \cite{CCN-traffic-optimization} which proposes a push mechanism for CCN on sensor networks.

Furthermore, the simplified communication model at the base of ICN was initially designed with the assumption that the number of consumers is much larger than the number of producers, targeting use cases that are comparable to the scenarios CDNs aim for. Such an assumption does not hold in general in the IoT, where consumers (e.g., a data sink) are often outnumbered by producers (e.g., sensors). In consequence, content caching strategies designed for scenarios similar to CDN will not be efficient in the IoT, and thus, alternative strategies should be designed for content replication and content cache replacement in the IoT with~ICN.

\section{Conclusion and Perspectives}
\label{sec:conclusion}

ICN has recently been identified as a potential alternative network paradigm for the Internet of Things. In this paper, we have carried out experiments with NDN on a real IoT deployment consisting in tens of constrained nodes in multiple rooms of multiple buildings. Based on this experience, we have shown that ICN is indeed applicable in the IoT, and that it can offer advantages over an approach based on 6LoWPAN/IPv6/RPL in terms of energy consumption, as well as in terms of RAM and ROM footprint. We have proposed several interoperable NDN enhancements to decrease  energy consumption and routing state. Furthermore, we identified several areas where future work is needed. Topics include (i) an efficient header compression and fragmentation/reassembly adaptation layer below NDN to fit typically small frame sizes, (ii) IoT-specific content replication and cache replacement strategies, (iii) enhancements of the basic ICN communication model to accommodate IoT traffic patterns, (iv) further studies on the impact of caching on content availability in the context of sleeping nodes, and (v) short naming schemes optimized for constrained devices.


 \smallskip
\noindent\textbf{Acknowledgments}\quad
We would like to thank the anonymous reviewers and our shepherd, Lan Wang,
for their valuable comments. Furthermore, we would like to thank Lixia
Zhang for first discussions about NDN support in RIOT. This work was
partially supported by ANR and BMBF within the SAFEST and Peeroskop projects,
and the DAAD within the guest lecture program. \ \\







\bibliographystyle{IEEEtran}
  \bibliography{bibliography,rfc}
\end{document}